# Machine Learning Inter-Atomic Potentials Generation Driven by Active Learning: A Case Study for Amorphous and Liquid Hafnium dioxide

Ganesh Sivaraman[1], Anand Narayanan Krishnamoorthy[2], Matthias Baur[2], Christian Holm[2], Marius Stan[3], Gábor Csányi[4], Chris Benmore[5], and Álvaro Vázquez-Mayagoitia[6*]

[1]*Leadership Computing Facility, Argonne National Laboratory, Lemont, 60439, IL, USA*
[2]*Institute for Computational Physics, Universität Stuttgart, Allmandring 3, 70569 Stuttgart, Germany*
[3]*Applied Materials Division, Argonne National Laboratory, Lemont, IL 60439, USA*
[4]*Department of Engineering, University of Cambridge, Trumpington Street, Cambridge, CB2 1PZ, United Kingdom*
[5]*X-ray Science Division, Argonne National Laboratory, Argonne, IL 60439, USA*
[6]*Computational Science Division, Argonne National Laboratory, Argonne, IL 60439, USA*

We propose a novel active learning scheme for automatically sampling a minimum number of uncorrelated configurations for fitting the Gaussian Approximation Potential (GAP). Our active learning scheme consists of an unsupervised machine learning (ML) scheme coupled to Bayesian optimization technique that evaluates the GAP model. We apply this scheme to a Hafnium dioxide ($HfO_2$) dataset generated from a 'melt-quench' *ab initio* molecular dynamics (AIMD) protocol. Our results show that the active learning scheme, with no prior knowledge of the dataset is able to extract a configuration that reaches the required energy fit tolerance. Further, molecular dynamics (MD) simulations performed using this active learned GAP model on 6144-atom systems of amorphous and liquid state elucidate the structural properties of $HfO_2$ with near *ab initio* precision and quench rates (i.e. 1.0 K/ps) not accessible via AIMD. The melt and amorphous x-ray structural factors generated from our simulation are in good agreement with experiment. Additionally, the calculated diffusion constants are in good agreement with previous *ab initio* studies.

---

[*]E-mail: vama@alcf.anl.gov

# 1 Introduction

Hafnium dioxide (HfO$_2$), or hafnia, is a relevant material in semiconductor process technology such as high-k gate dielectrics[i-ii] could be excellent replacement for silicon dioxide (SiO$_2$) also used in switching mechanisms for resistive random-access memories (RRAMs). Furthermore, hafnia is used as high temperature refractory material[iii] and also has applications in nuclear technologies. [iv] In particular, high-k gate dielectrics applications require thin films of amorphous HfO$_2$[v] (a-HfO$_2$). Thermal atomic layer deposition grown thin films of hafnia on silicon wafers has been shown to exhibit amorphous structures with embedded crystallites.[vi] Experimental measurements performed on an agglomeration of low-density a-HfO$_2$ nanoparticles in comparison with molecular dynamics (MD) simulations have revealed the presence of monoclinic like embedded crystallites.[vii] The density of the a-HfO$_2$ is also shown to significantly influence the atomistic structure and oxygen diffusion. [vii-viii] In addition, there are well known variability of the Hf-O coordination number (between 6.1 to 7.6) reported in MD simulation of a-HfO$_2$, a detailed survey of which is available in this previous study. [vii] Hence it is critical to characterize and tune the local atomic structure of such application relevant amorphous materials. Subsequent studies have shown that the investigated atomic packing arrangements in amorphous hafnia is highly dependent on the density of the material. [vii-viii]

MD simulations based on Density Functional Theory (DFT) can provide structural description of amorphous materials at quantum mechanical accuracy. But such calculations are severely limited by the finite system size (10-100's of atoms) and short time scales (~ 10's of ps). On the other hand, traditional MD simulations based on inter atomic potentials derived from empirical and physical approximations on the other hand can provide access to larger system size (1000's of atoms) with longer time scales (~ 100's of ps) by sacrificing the quantum mechanical accuracy. Inverse modeling techniques such as Reverse Monte Carlo (RMC),[ix] along with advanced experimental techniques such as synchrotron based high-energy x-ray diffraction has certainly aided in improved understanding of the atomic structure of amorphous materials, but such techniques can only provide a statistical description of the local atomic environment.[x]

In the age of 'big-data' driven materials informatics,[xi] emerged a new generation of machine learning (ML) inter-atomic potentials.[xii,xiii,xiv,xvi,xvii,xviii,xix,xx] Unlike classical inter atomic potentials, these potentials use ML methods to map the direct functional relationship between atomic configuration and energy from reference quantum mechanical calculated datasets. Subsequently, many recent applications of ML inter-atomic potentials have achieved simulation length and time scale accessible to classical inter-atomic potentials, with near quantum mechanical accuracies.[xxi-xxii] These advances have certainly impacted the amorphous materials modeling where in a recent study,[xxiii] a realistic structure of amorphous silicon was achieved based on the Gaussian Approximation potential (GAP) framework.[xxiv] In that study, a 4096-atom simulation system was melted, and slow cooled over a 10ns quench simulation. The first sharp diffraction peak (fsdp) magnitude in the structure factor for that system was found to have the closest agreement to experiment for amorphous silicon. Another study showcased the modeling of a more complex ternary phase-change memory material ($Ge_2Sb_2Te_5$) using the GAP framework.[xxv]

Encouraged by these results, we construct GAP model to develop a potential for amorphous (a-$HfO_2$) and liquid hafnia (l-$HfO_2$), details of which are discussed in the methods section. The reference dataset would be created from AIMD melt-quench protocol. We also propose an active learning scheme to automatically sample configurations from the AIMD generated dataset that would lead to a near DFT precision on the model predicted energies. The aim of using this approach is to systematically reduce the number samples for training assuring predictability within an arbitrary mean absolute error, which later produces a reduced size model that requires less computational resources to train and evaluate. Currently machine learned atomistic models bridges the gap between DFT and classical methods by showing *ab initio* accuracy for several thousands of atoms with linear scaling. Thus providing much accurate insights into bulk properties of liquids, amorphous and solid state systems [xxiii,xxv,xxvi] and comparison to experimental observables serves as a benchmark to train the acceptable machine learned based potentials. Furthermore, we stress the fact that all the current models are trained only on the *ab inito* data and experimental entities are used for benchmarking purposes for improving the training dataset. Here we discuss the application of the active learning to automate the training process of these machine learned atomistic models. Understanding the process and application of these models using amorphous and liquid Hafnia system is achieved through a discussion on the validation of the active learned total potential with theoretical and experimental data. The effect of quench rate is investigated via the order parameters extracted from the MD simulation of a medium size system (768 atom). Finally, we have performed a melt-quench MD on a large system size (6144 atoms) to generate a-$HfO_2$, by utilizing the active learned GAP model. The results are validated against X-ray diffraction measurements.

## 2   Results and discussion

**Active learning**

The optimal training configurations are generated via the active learning workflow described in the methods section. The configuration for building up the potentials where chosen with active learning workflow in order to achieve standard error convergence pertaining to the range of properties measured. [xxiii] The validation plot for the active learned a-HfO$_2$ potential is shown in Figure 1. We start by discussing the inset table of Figure 1, where the details of the active learned training configurations have been summarized. The number of iterations ($N_{iter}$) to reach this target convergence and the resulting number of training configuration ($N_{train}$). The energy convergence criterion, $E_{tol}$ was set to 5 meV/atom for quenching dataset, 2 meV/atom for the liquid and amorphous respectively. In the case of amorphous and liquid phases, the workflow ended up with configurations with relative smaller iterations. It can be observed that the non-equilibrium nature of the quenching procedure over a large temperature range leads to significant challenge in picking the right training configuration. Consequently, it took the workflow 11 iterations to reach the requested accuracy. But $N_{train}$ = 260 is a meager ~0.8% of the entire AIMD quench dataset. This would be a significant human effort if done by hand picking configurations from the *ab initio* data set. The human choice of training data set is based out of previous experience and literature reviews, combined with trial and error principle to achieve the desired error convergence. However, for the system relevant to this study, the active learning workflow gives an automated path to achieve the desired accuracy without human intervention. Interested readers are advised to refer to the Supplementary Information (SI) for a discussion on manual configuration selection and its benchmark with respect to the active learning scheme presented here.

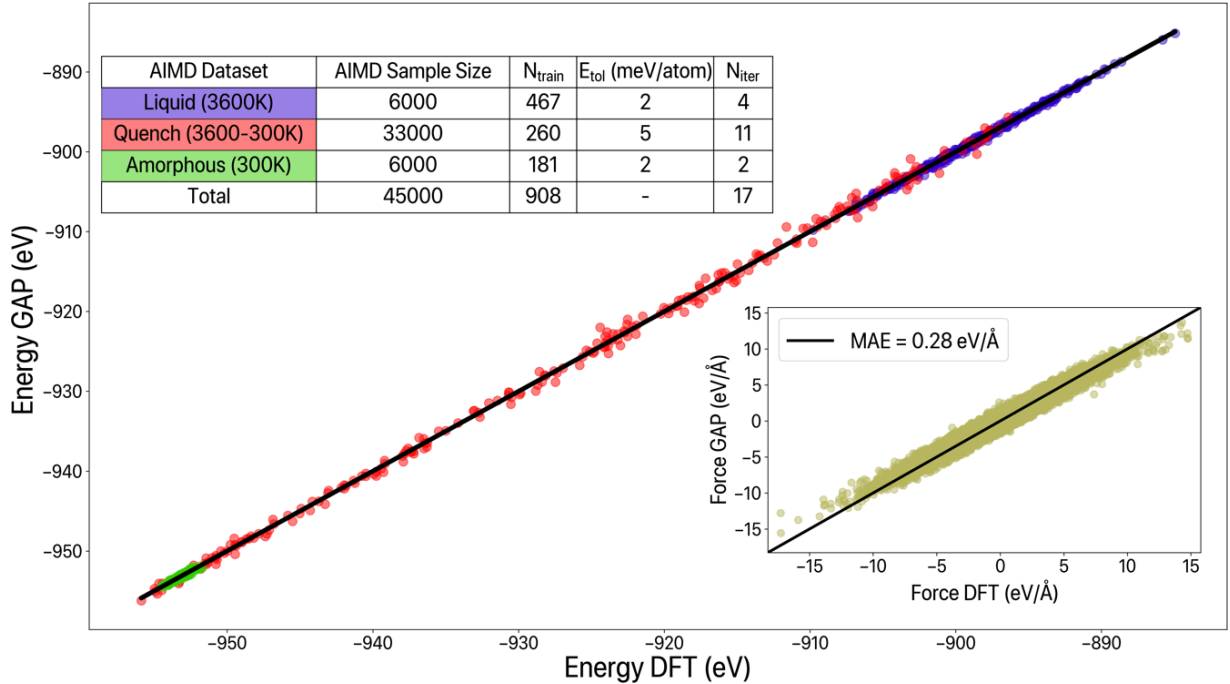

Figure 1: The GAP predicted vs DFT energy validation plot for the active learned a-HfO$_2$ potential. The validation was performed on a test dataset independent from the training data. The scatter color indicates the AIMD dataset source from which the test data point was chosen. (Inset table) Summary of the AIMD datasets, and active learning settings. (Inset plot) Force validation plot.

The individual active learned configuration was combined to train the final a-HfO$_2$ potential. The details of the hyper parameters are available in the SI. Now we turn our attention to the quality of this active learning a-HfO$_2$ potential, by validating an independent test dataset. It can be seen that the active learned potential gives close to linear fit in predicted GAP energies vs DFT energies. The overall Mean Absolute Error (MAE) for GAP predicted energy is 2.6 meV/atom. In the inset, we also show force convergence with an overall MAE of 0.28 eV/Ang. This supports the argument that GAP based atomistic models predict the local
6

properties of the systems with good accuracy of MAE < 5 meV/atom for liquid and amorphous states.[xxvii,xxviii]

## Active Learned GAP MD

With the active learned GAP potentials, MD simulations were performed using the LAMMPS package.[xlix] We obtain the amorphous structures from standard melt-quench scheme as described in the methods section. The details of the melt-quench are described in this section.

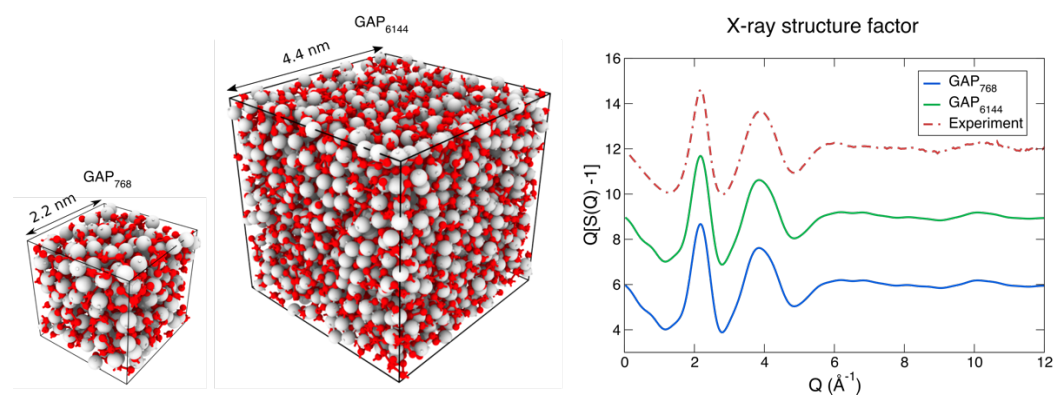

Figure 2: (Left panel) The simulation set up for GAP MD. A 768-atom cell was generated by 2x2x2 replication of a random snapshot extracted from the AIMD liquid dataset. Another set of 6144 atom simulation cell was generated by 4x4x4 replication of a Packmol generated 96 atom configurations. Hf (silver), Oxygen (red). (Right panel) Comparison of X-ray structure factors for l-$HfO_2$ with simulated structure factors obtained from GAP-MD.

Figure 2 and Figure 3 represent the structure factor of l-$HfO_2$ and a-$HfO_2$ respectively. The X-

ray structure factor of molten hafnia at 3173.15 K (2900 °C) was measured out to a Q-value of 22.5 Å$^{-1}$. The simulated atom-atom partial X-ray structure factors are obtained via inverse Fourier transform of pair distribution functions, weighted by the appropriate (Q-space) x-ray form and concentration factors, summed and compared directly with the experimental data. We can see a very good agreement of our atomistic model structure factor to that of the experimental X-ray diffraction experiments for both l-HfO$_2$ and a-HfO$_2$ as reported in Figure 2 and 3 respectively. The automated atomistic modeling with active learning techniques is capable of capturing the salient structural features upon changing from an equilibrium liquid structure to a non-equilibrium amorphous state. To highlight the detailed structural rearrangements between the liquid and amorphous structures we plot Q[S(Q)-1] to emphasize the strong oscillations in amorphous signal in the range Q~5-15 Å$^{-1}$, which are not present in the liquid signal. As expected, the oscillations decays for Q>5 Å$^{-1}$ in the liquid structure factor due to the increased local disorder at higher temperature.

The other major advantage of our model is that it is parameter free and it can be improved with the addition of data sets to the existing potential training data. Here we have a-HfO$_2$ potential that can mimic both liquid and amorphous states to *ab initio* accuracies further exploring temperature range of the whole quench domain as trained with large datasets. This can be obtained in an automated workflow using active learning without human intervention**.** We direct the readers to refer to the methods section and the recent success in the other applications of active learning.[xxx,xxxi,xxxii]

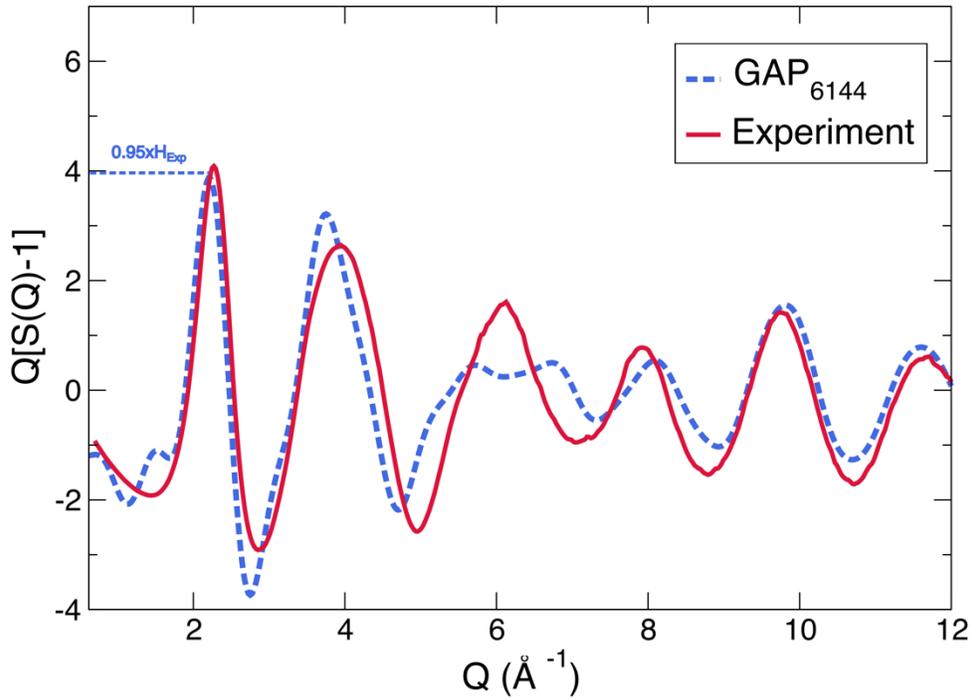

Figure 3: Comparison of experiment X-ray structure factors for a-HfO2 (red line) with simulated structure factors obtained from GAP-MD (Blue dots).

The objective of using our atomistic model from machine learning GAP model is to attain *ab initio* accuracies with scaling which cannot be accessed with DFT. As described in the Methods section, the *ab initio* reference data has 96 atom system and our classical MD simulation consists of system with 6144 atoms in total with box size of 4.4×4.4×4.4 nm$^3$. From Figure 2, it is seen that with increase in scale the accuracy of structure factor of liquid hafnia is already captured for the 6144 atoms system, thus maintaining the DFT-quality MD. This supports the argument that our GAP based atomistic model can retain DFT accuracies at large scales with increased simulation times with linear scaling.[xxiii]

To characterize the atomic structure in finer detail we show the partial Pair Distribution Functions (PDFs) from GAP MD simulations for both l-HfO$_2$ and a-HfO$_2$ in Figure 4.

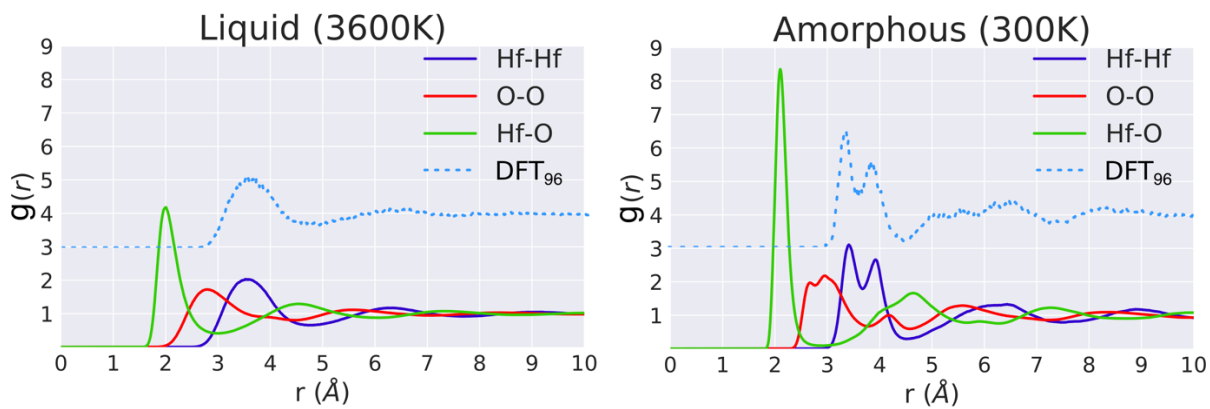

Figure 4: The partial radial distribution functions for (left panel) l-HfO$_2$, and (right panel) a-HfO$_2$ derived from the GAP MD model in a 6144-atom cell. The dotted line shows the baseline Hf-Hf PDFs derived from DFT 96 atom cell.

The calculated partial PDF's illustrate the growth of intermediate range ordering in a-HfO$_2$ compared to liquid in real space, see Figure 4. The first peak at ~2 Å correspond to the average bond length between hafnium and oxygen. For l-HfO$_2$, there are single broad peaks associated with the Hf-O and Hf-Hf correlations, but for a-HfO$_2$ the Hf-O peak becomes narrower and increases in intensity. Moreover, the broad Hf-Hf peak in the liquid splits into two peaks in the amorphous form, corresponding to well defined edge-sharing polyhedra at 3.4 Å and corner-sharing polyherdra 3.9 Å. The ratio of the edge/corner sharing ratio is known to be density dependent[viii] and leads to the formation of a disordered network at distances r > 8 Å in the amorphous phase, which have also been observed in previous *ab initio* studies and experiments.[vii] The light blue dotted lines in Figure 4 represent the partial PDFs obtained from ab-initio MD simulations and our GAP MD model for 6144 atoms accurately reproduces the Hf-Hf peak split corresponding to the edge-sharing and corner-sharing polyhedra seen in the baseline DFT.

| Method | Hf-O (CN) | Hf-O peak(A) | Hf-Hf peak (A) | Density (g cm$^{-3}$) |
|---|---|---|---|---|
| Exp a-HfO2 | 6.8±0.6 | **2.13** | 3.38(1), 3.89(1) | 7.69 |
| GAP a-HfO2 | 6.6 | **2.12** | 3.41,3.92 | 7.69 |
| Broglia a-HfO2 | 6.2 | 2.15 | 3.37, 3.92 | 7.69 |
| Exp l-HfO2 | 7.0±0.6 | 2.05 | 3.67 | 8.16 |
| GAP l-HfO2 | 6.13 | 2.00 | 3.59 | 8.16 |

Table1: Local structure properties extracted from experiment,[vii] a classical force field (Broglia *et. al.*[viii] ) and GAP MD (this work).

Table 1 gives the coordination numbers and bond lengths of a-HfO$_2$ and l-HfO$_2$ in comparison to experiment. GAP MD gives a very close agreement with the Hf-O coordination number and bond lengths with experiments. From Table 1, it can also be seen that hafnium gives a seven-fold coordination with oxygen, which is similar to the monoclinic phase asymmetric arrangements of Hf-O bond distances at 2.03 - 2.25 Å as reported in previous study. [xxxv] To enunciate the argument from the above, we calculated the bond angle distribution of Hf-O-Hf of a-HfO$_2$ the results of which are discussed below.

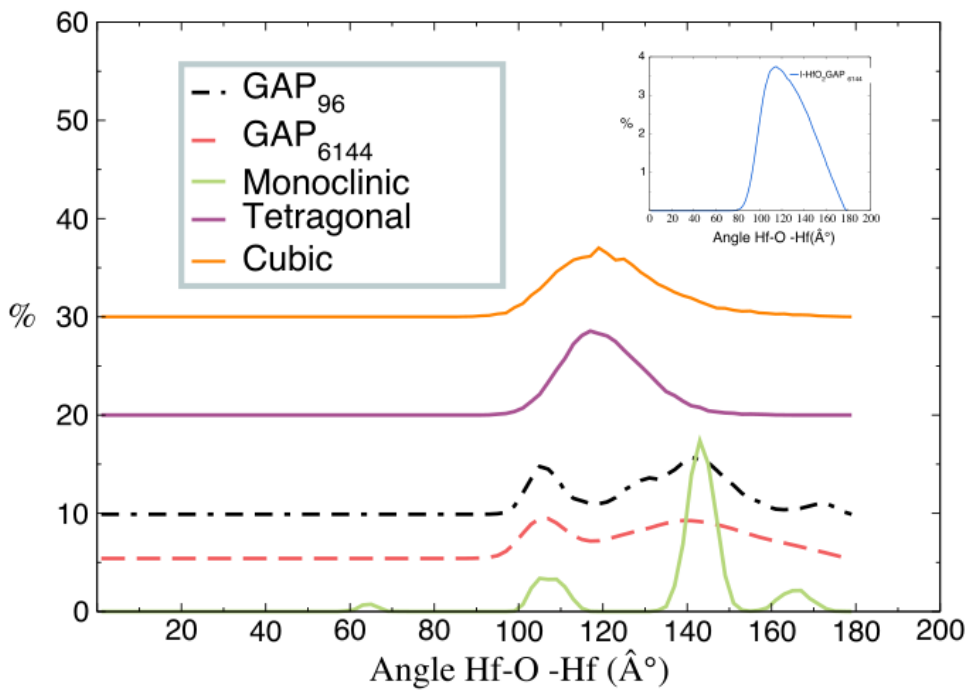

Figure 5: Bond angle distribution obtained from AIMD of pure phases of HfO$_2$. GAP predicted bond angle distributions for a-HfO$_2$. The distributions have been shifted in vertical axes for clarity. The inset plots show the bond angle distribution derived from the GAP model for l-HfO$_2$.

The bond angle distributions derived from the GAP-MD are shown in Figure 5. For a-HfO$_2$, it can be seen that there are two peaks at Hf-O-Hf angles of 107° and 142°. Previous studies have attributed these peaks to the existence of edge and corner sharing polyhedra,[viii] respectively. In the GAP models the two Hf-O-Hf peaks suggests a structural morphology in a-HfO$_2$ resembling that of the monoclinic phase, rather than a single broad peak around 120° observed in the tetragonal and cubic phases. However, a major difference between the amorphous Hf-O-Hf bond angle distribution and that of the crystalline monoclinc form, is the width and intensity of the 142° peak. The broad nature of this peak in a-HfO$_2$ is indicative of a wide distribution of packing arrangements of corner shared HfO$_n$ polyhedra compared to m-HfO$_2$. The similarity of the 107° Hf-O-Hf bond angle peak between a-HfO$_2$ and m-HfO$_2$ can be understood due to the strict geometric requirements of edge shared units. Previous *ab initio* studies have predicted two different types of amorphous structure formation using the melt quench scheme to investigate the amorphous to crystalline phase transition. They related these structures to tetragonal type and monoclinic type with respect to their long-range ordering and volume energy curve.[xxxv] The l-HfO$_2$, on the other hand has a single asymmetric Hf-O-Hf bond angle distribution peak located at ~ 115°, similar to that of the cubic and tetragonal phase of hafnia which have bond angle distribution peaks in the interval of ~117-120°.

In summary, we have validated GAP based atomistic models using structural properties. Furthermore, we also note that the diffusive behavior of amorphous and liquid hafnia has been studied previously for their dielectric properties in RRAM.[viii,vii] In l-HfO$_2$ at 3100 K, we estimate hafnium and oxygen diffusion constants to be {Hf: 3.3796 (± 0.01), O: 6.2971 (±0.01)} × 10$^{-5}$ cm$^2$/s respectively. This is in good agreement with the previous *ab inito* study performed by Hong *et. al.*[xxxvi] reported at 3100K {Hf: 2.4 (±0.1), O:5.2 (±0.2)} × 10$^{-5}$ cm$^2$/s. As reported in the Table I of Hong *et. al.,* the computational cost associated with that *ab inito* calculation to achieve 56ps of trajectory for a system size of 270 atom is 21500 cpu hours. Our GAP model with 6144 atom unit cell on the other hand achieved 110ps with a computational cost of 1080 cpu hours, while retaining the *ab inito* accuracy.

**Conclusion**

To conclude, we showed that data driven techniques can improve the quality of interatomic potentials used for MD simulations up-to an accuracy of *ab inito* training data. Further we use an active learning scheme to automate the process of choosing optimal training and test configurations from the available data sets to attain desired convergence. We do understand that there will be scenarios where human logic and interventions are required to guide the training process. However, for the system relevant to this study, the active learning workflow gives an automated path to achieve the desired accuracy without human intervention. We showed here the proof of concept that the active learning could be a possible replacement to educated hand-picking configuration method. Thus, enhancing the speed of the learning process to achieve the desired atomistic interatomic potentials. Here we used Gaussian Approximation Potential (GAP) based atomistic models to model a-$HfO_2$ to showcase the active learning workflow. Our machine learned atomistic model showed very good agreement with experimental liquid and amorphous x-ray structure factors. Our model is able to predict the diffusion constants at the same order of magnitude as that of previous study,[xxxvi] due to the linear scaling of GAP models. We also exemplify the fact that the accuracy of the atomistic potential purely depends on the quality of quantum mechanical calculations used for training. This method can further be used to enhance the speed of modelling amorphous and liquid systems of interest.

## 3 Methods

### AIMD Dataset

The input for amorphous structures was generated using Packmol[xxxvii] by packing 96 random particles (32 Hf, 64 O) in a cubic box of density 8.16 g/cm$^{-3}$.[xxxviii] The density of this cubic box was taken to be experimental density reported by Gallington *et. al.*[vii] The initial configuration was heated to 3600K (500K higher than melting point) and sampled for 12 ps. The final snapshot from the liquid configuration at 3600K is quenched to 300K at a rate of 100K/ps. The final configuration of quench was rescaled to experimental density of 7.69 g/cm$^{-3}$ and 12 ps of amorphous configurations were generated. From the liquid and amorphous trajectories, the final 6ps of snapshots were retained in the dataset. All of the 33ps of quenching trajectories were retained in the dataset. These same density values are used through this study.

The atomic configurations, energy, force, and virial stress were calculated in NVT ensemble using the Vienna Ab initio Simulation Package (VASP).[xxxix-xl] A plane wave cut off of 520 eV (30% larger than the largest cut off value), 2x2x2 K-grid Monkhorst-Pack scheme, and 1fs time step were used. The Perdew-Burke-Ernzerhof exchange-correlation functional[xli] and projector-augmented wave method[xlii] were employed. Further details are available in the Supplementary Information.

**Active learning**

Our aim is to employ active learning to automatically sample uncorrelated learning configurations from the reference datasets, exploiting the underlying cluster structure embedded in the dataset[xliii] and partitions them in to 'N' uncorrelated clusters[xliv]. To this end, we used HBSCAN[xlv] a non-parametric unsupervised learning method. This algorithm has been successfully applied in partitioning and analyzing molecular dynamics trajectory[xlvi]. The individual AIMD trajectory is passed to the active learning workflow (Figure 2). In the first step, HDBSCAN partitioned the input trajectories in to uncorrelated clusters. Once the information on the clusters is extracted, a series of trials are run to sample data from these clusters. In each trial, sample are drawn from each cluster at intervals separate by K, where K goes from $K_{max}$ to $K_{min}/2$. $K_{max}$ is the sample size of the largest cluster. In the initial trial, exactly one unique training and test configuration would be sampled from each cluster (i.e. sampling width is same as maximum sample size, $K_{max}$). At this point a Bayesian optimization (BO) method [xlvii] is invoked which runs a series of convergence checks by generating GAP models from training data and validating them on test set of configurations. If the best GAP model generated by BO does not achieve the required accuracy, then another trial is run where more data is sampled from the largest clusters based on the sampling criterion explained above. The workflow stops the iterations ($N_{iter}$) as soon as the Mean Absolute Error (MAE) in energy is on or below a specified threshold tolerance value ($E_{tol}$). For the sake of consistency, an equal number of unique training ($N_{train}$) and test configurations were sampled from each cluster per trial. The optimal configuration extracted from each of the individual AIMD datasets are sequentially combined to train the a-$HfO_2$ potential. The cluster parameters such as N, $K_{min}$, and $K_{max}$ are outcomes of the unsupervised learning, consequently the sampling regime is expected to generalize easily to new dataset and requires no hand tuning.

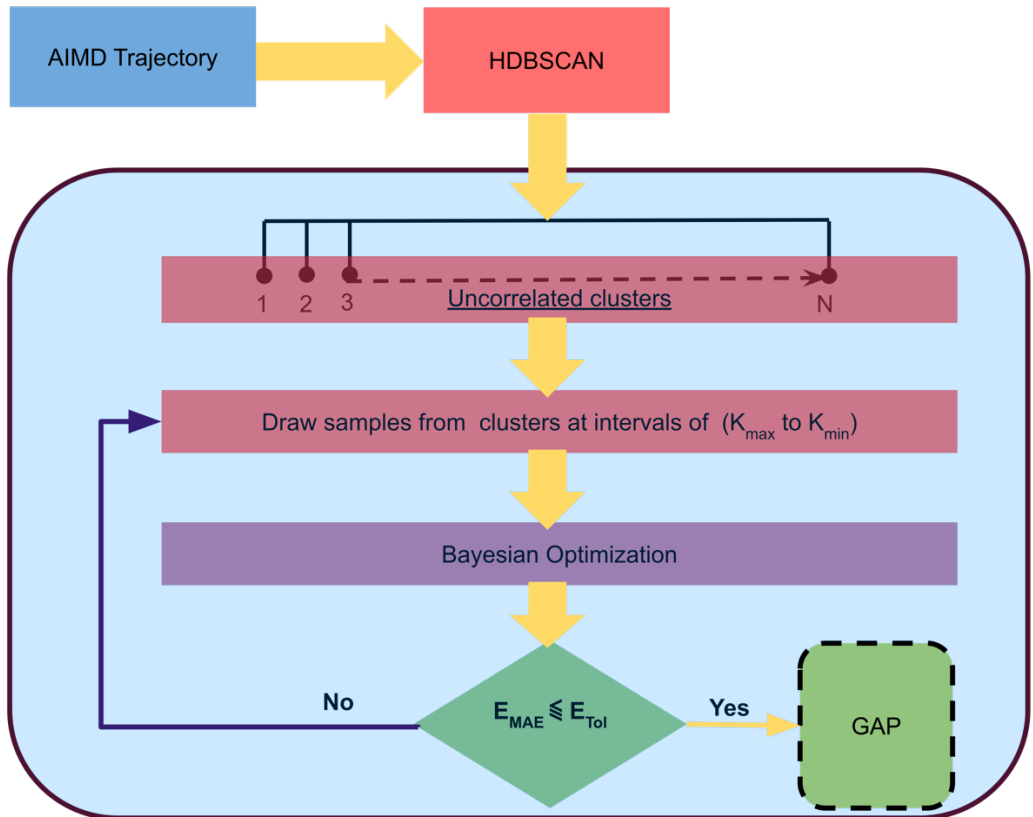

Figure 6: Active learning workflow

**GAP MD**

The individual active learned datasets were combined to train the total gap potential. In order to prevent unphysical clustering of oxygen atoms at high temperature simulations, a non-parametric two body was added to the Smooth Overlap of Atomic Position (SOAP) descriptor.[xlviii] The details of the training hyper parameters are provided in the supplementary information. The total gap potential is used to perform melt-quench MD simulation using the LAMMPS molecular dynamics package.[xlix] Two different system sizes were considered for the simulations. A NVT ensemble was used to sample the configurations with a time step of 1 fs. The liquid and amorphous configurations were samples for 100 - 200 ps. Different quench rates and system sizes were also investigated details of which are discussed in the results section.

**Experiment**

High energy x-ray diffraction experiments on liquid and amorphous $HfO_2$ were performed on beamline 6-ID-D at the Advanced Photon Source. The details have previously been reported elsewhere[vii], so only the salient information is provided here. The liquid state diffraction measurements were performed by laser heating in an aerodynamic levitator using 100.27 keV x-rays, in combination with a large a-Si area detector. The experiments on amorphous $HfO_2$ were performed in a similar manner but using an incident energy of 131.74 keV to attain data out to higher Q-values and improve the real space resolution. The data were analyzed using standard correction procedures, including corrections for background, Compton scattering, fluorescence and oblique incidence to yield the Faber-Ziman total x-ray structure factors. A Sine Fourier transform was used to obtain the corresponding x-ray pair distribution functions (PDFs).

## 4 Ethics declarations
**Competing interests**
The authors declare no competing interests.

## 5 Acknowledgements

This material is based upon work supported by Laboratory Directed Research and Development (LDRD) funding from Argonne National Laboratory, provided by the Director, Office of Science, of the U.S. Department of Energy under Contract No. DE-AC02-06CH11357. Argonne National Laboratory's work was supported by the U.S. Department of Energy, Office of Science, under contract DE-AC02-06CH11357. We gratefully acknowledge the computing resources provided on Bebop; a high-performance computing cluster operated by the Laboratory Computing Resource Center at Argonne National Laboratory. This research used resources of the Advanced Photon Source, a U.S. Department of Energy (DOE) Office of Science User Facility operated for the DOE Office of Science by Argonne National Laboratory under Contract No. DE-AC02-06CH11357. Use of the Center for Nanoscale Materials, an Office of Science user facility, was supported by the U.S. Department of Energy, Office of Science, Office of Basic Energy Sciences, under Contract No. DE-AC02-06CH11357. A.N.K gratefully acknowledges useful discussions with Dr. Jens Smiatek, Dr. Frank Uhlig and financial support from the German Funding Agency (Deutsche Forschungsgemeinschaft-DFG) through the SimTech Cluster of Excellence (EXC 310).

# 6 Authors' contributions

C.B., M.S., and A.V.M. jointly conceived the problem statement. C.B performed the experimental measurements. G.S. implemented the active learning workflow, computed the dataset, executed the active learning driven GAP fitting, and performed the GAP-MD simulations with input from C.B., A.V. M and G. C.

A.N.K and M.B. performed the hand picking of configurations and GAP benchmarks with inputs from C. H.

G.S. and A.N.K performed the data analysis with inputs from C.B, G. C., and A.V.M.

G.S., A.N.K, C.B., and A.V.M. wrote the paper with input from all authors. All authors revised the paper and approved its final version.

[i] Schlom, D. G., Guha, S., & Datta, S. Gate oxides beyond SiO$_2$. *MRS bulletin*, **33(11)**, 1017-1025 (2008).

[ii] Matthews, J. N. A. Semiconductor industry switches to hafnium-based transistors. Physics Today **61**, 2, 25 (2008)

[iii] Upadhya, K., Yang, J. M., & Hoffman, W. P. Advanced materials for ultrahigh temperature structural applications above 2000° C. *Am. Ceram. Soc. Bull*, **76(12),** 51-56 (1997).

[iv] Wang, J., Li, H. P., & Stevens, R. Hafnia and hafnia-toughened ceramics. *Journal of materials science*, **27(20)**, 5397-5430 (1992).

[v] Li, F.M., Bayer, B.C., Hofmann, S., Dutson, J.D., Wakeham, S.J., Thwaites, M.J., Milne, W.I. & Flewitt, A.J. High-k (k= 30) amorphous hafnium oxide films from high rate room temperature deposition. *Applied Physics Letters*, **98(25)**, p.252903 (2011).

[vi] Miranda, A. *Understanding the Structure of Amorphous Thin Film Hafnia-Final Paper* (No. **SLAC-TN-15-066**). SLAC National Accelerator Lab., Menlo Park, CA (United States) (2015).

[vii] Gallington, L., Ghadar, Y., Skinner, L., Weber, J., Ushakov, S., Navrotsky, A., Vazquez-Mayagoitia, A., Neuefeind, J., Stan, M., Low, J. & Benmore, C. The structure of liquid and amorphous hafnia. *Materials*, **10(11)**, 1290 (2017).

[viii] Broglia, G., Ori, G., Larcher, L., & Montorsi, M. Molecular dynamics simulation of amorphous HfO2 for resistive RAM applications. *Modelling and Simulation in Materials Science and Engineering*, **22(6)**, 065006. (2014)

[ix] McGreevy, R. L., & Pusztai, L. Reverse Monte Carlo simulation: a new technique for the determination of disordered structures. *Molecular simulation*, **1(6)**, 359-367 (1988).

[x] McGreevy, R. L. Reverse monte carlo modelling. *Journal of Physics: Condensed Matter*, **13(46)**, R877 (2001).

[xi] Jain, A., Hautier, G., Ong, S. P., & Persson, K. New opportunities for materials informatics: Resources and data mining techniques for uncovering hidden relationships. *Journal of Materials Research*, **31(8)**, 977-994 (2016).

[xii] Behler, J. Perspective: Machine learning potentials for atomistic simulations. *The Journal of chemical physics*, **145(17)**, 170901 (2016).

[xiii] Behler, J., & Parrinello, M. Generalized neural-network representation of high-dimensional potential-energy surfaces. *Physical review letters*, **98(14)**, 146401 (2007).

[xiv] Thompson, A. P., Swiler, L. P., Trott, C. R., Foiles, S. M., & Tucker, G. J. Spectral neighbor analysis method for automated generation of quantum-accurate interatomic potentials. *Journal of Computational Physics*, **285**, 316-330 (2015).

[xvi] Shapeev, A. V. Moment tensor potentials: A class of systematically improvable interatomic potentials. *Multiscale Modeling & Simulation*, **14(3)**, 1153-1173 (2016).

[xvii] Smith, J. S., Isayev, O., & Roitberg, A. E. ANI-1: an extensible neural network potential with DFT accuracy at force field computational cost. *Chemical science*, **8(4)**, 3192-3203 (2017).

[xviii] Huan, T. D., Batra, R., Chapman, J., Krishnan, S., Chen, L., & Ramprasad, R. A universal strategy for the creation of machine learning-based atomistic force fields. *NPJ Computational Materials*, **3(1)**, 37 (2017).

[xix] Li, Z., Kermode, J. R., & De Vita, A. Molecular dynamics with on-the-fly machine learning of quantum-mechanical forces. *Physical review letters*, **114(9)**, 096405 (2015).

[xx] Schütt, K. T., Sauceda, H. E., Kindermans, P. J., Tkatchenko, A., & Müller, K. R. SchNet–A deep learning architecture for molecules and materials. *The Journal of Chemical Physics*, **148(24)**, 241722 (2018).

[xxi] Chmiela, S., Sauceda, H. E., Poltavsky, I., Müller, K. R., & Tkatchenko, A. sGDML: Constructing accurate and

data efficient molecular force fields using machine learning. *Computer Physics Communications (2019)*.

Zuo, Y., Chen, C., Li, X., Deng, Z., Chen, Y., Behler, J., Csányi, G., Shapeev, A.V., Thompson, A.P., Wood, M.A. & Ong, S.P. A Performance and Cost Assessment of Machine Learning Interatomic Potentials. *arXiv preprint arXiv:1906.08888 (2019)*.

[xxiii] Deringer, V. L., Bernstein, N., Bartók, A. P., Cliffe, M. J., Kerber, R. N., Marbella, L. E., ... & Csányi, G. Realistic atomistic structure of amorphous silicon from machine-learning-driven molecular dynamics. *The journal of physical chemistry letters*, **9(11)**, 2879-2885 (2018).

[xxiv] Bartók, A. P., Payne, M. C., Kondor, R., & Csányi, G. Gaussian approximation potentials: The accuracy of quantum mechanics, without the electrons. *Physical review letters*, **104(13)**, 136403 (2010).

[xxv] Mocanu, F. C., Konstantinou, K., Lee, T. H., Bernstein, N., Deringer, V. L., Csányi, G., & Elliott, S. R. Modeling the Phase-Change Memory Material, Ge2Sb2Te5, with a Machine-Learned Interatomic Potential. *The Journal of Physical Chemistry B*, **122(38)**, 8998-9006 (2018).

[xxvi] Morawietz, T., Singraber, A., Dellago, C., & Behler, J. How van der Waals interactions determine the unique properties of water. *Proceedings of the National Academy of Sciences*, **113(30)**, 8368-8373 (2016).

[xxvii] Deringer, V. L., Pickard, C. J., & Csányi, G. Data-driven learning of total and local energies in elemental boron. *Physical review letters*, **120(15)**, 156001 (2018).

[xxviii] Bartók, A. P., Kermode, J., Bernstein, N., & Csányi, G. Machine learning a general-purpose interatomic potential for silicon. *Physical Review X*, **8(4)**, 041048 (2018).

[xxx] Gubaev, K., Podryabinkin, E. V., Hart, G. L., & Shapeev, A. V. Accelerating high-throughput searches for new alloys with active learning of interatomic potentials. *Computational Materials Science*, **156**, 148-156 (2019).

[xxxi] Bernstein, N., Csányi, G., & Deringer, V. L. De novo exploration and self-guided learning of potential-energy surfaces. *arXiv preprint arXiv:1905.10407* (2019).

[xxxii] Zhang, L., Lin, D. Y., Wang, H., Car, R., & Weinan, E. (2019). Active learning of uniformly accurate interatomic potentials for materials simulation. *Physical Review Materials*, **3(2)**, 023804.

[xxxv] Luo, X., & Demkov, A. A. Structure, thermodynamics, and crystallization of amorphous hafnia. *Journal of Applied Physics*, **118(12)**, 124105 (2015).

[xxxvi] Hong, Q. J., Ushakov, S. V., Kapush, D., Benmore, C. J., Weber, R. J., van de Walle, A., & Navrotsky, A. Combined computational and experimental investigation of high temperature thermodynamics and structure of cubic ZrO 2 and HfO 2. *Scientific reports*, **8(1)**, 14962 (2018).

[xxxvii] Martínez, L., Andrade, R., Birgin, E. G., & Martínez, J. M. (2009). PACKMOL: a package for building initial configurations for molecular dynamics simulations. *Journal of computational chemistry*, **30(13)**, 2157-2164.

[xxxviii] Aykol, M., Dwaraknath, S. S., Sun, W., & Persson, K. A. Thermodynamic limit for synthesis of metastable inorganic materials. *Science advances*, **4(4)**, eaaq0148 **(2018)**.

[xxxix] Kresse, G., & Furthmüller, J. Efficiency of ab-initio total energy calculations for metals and semiconductors using a plane-wave basis set. *Computational materials science*, **6(1)**, 15-50 (1996).

[xl] Kresse, G., & Furthmüller, J. (1996). Software VASP, vienna . *Phys. Rev. B*, **54(11)**, 169 (1999).

[xli] Perdew, J. P., Burke, K., & Ernzerhof, M. Generalized gradient approximation made simple. *Physical review letters*, **77(18)**, 3865 (1996).

[xlii] Kresse, G., & Joubert, D. From ultrasoft pseudopotentials to the projector augmented-wave method. *Physical*

*Review B*, **59(3)**, 1758 (1999).

[xliii] Dasgupta, S., & Hsu, D. Hierarchical sampling for active learning. In *Proceedings of the 25th international conference on Machine learning* (pp. 208-215). ACM (2008).

[xliv] Dasgupta, S. Two faces of active learning. *Theoretical computer science*, **412(19)**, 1767-1781 (2011).

[xlv] McInnes, L., Healy, J., & Astels, S. hdbscan: Hierarchical density-based clustering. *J. Open Source Software*, **2(11)**, 205 (2017).

[xlvi] Melvin, R. L., Godwin, R. C., Xiao, J., Thompson, W. G., Berenhaut, K. S., & Salsbury Jr, F. R. Uncovering large-scale conformational change in molecular dynamics without prior knowledge. *Journal of chemical theory and computation*, **12(12)**, 6130-6146 (2016).

[xlvii] González, J. GPyOpt: a Bayesian optimization framework in Python. (2016).

[xlviii] Bartók, A. P., Kondor, R., & Csányi, G. On representing chemical environments. *Physical Review B*, **87(18)**, 184115 (2013).

[xlix] Plimpton, S. Fast parallel algorithms for short-range molecular dynamics. *Journal of computational physics*, **117(1)**, 1-19 (1995).